\def\L{{\cal L}}
\def\Lb{{\Lambda}}
\def\A{{\cal A}}
\def\E{{\cal E}}
\def\I{{\cal I}}
\def\J{{\cal J}}
\documentstyle[12pt]{article}
\renewcommand{\baselinestretch}{1.25}

\begin{document}
\begin{flushright}
DO-TH-94/10 \\
(May 1994)
\end{flushright}
\vspace{1.8cm}
\centerline{\Large \bf Aspects of Standard Models with Two Higgs
Doublets~\footnote{{\normalsize talk presented at the {\it XXIXth Rencontres
de Moriond (Electroweak Interactions and Unified Theories)}, M\'eribel,
France, March 12-19, 1994}}}
\vspace{0.8cm}
\begin{center}
{{\bf G.~Cveti\v c} \\
Inst.~f\"ur Physik, Universit\"at Dortmund, 44221 Dortmund, Germany}
\end{center}
\vspace{6.cm}
{\noindent {\bf ABSTRACT}}

\renewcommand{\baselinestretch}{0.9}
\baselineskip0.6ex


We~\footnote{{\normalsize works done in collaboration with {\bf C.S.~Kim},
Yonsei Univ., Seoul, Korea}}
present some properties of the SM with two Higgs doublets. Unlike
the minimal SM, the Yukawa couplings (in the usual ``type II''
model) converge with the increasing energy to flavor democracy (FD),
i.e.~to common values, in a specific flavor basis. This may represent
a possible signal of some new, (almost) flavor-blind physics beyond
the SM. When imposing the assumption of equality of the corresponding
quark and leptonic Yukawa couplings at high transition energies, we
can estimate the physical mass of the tau-neutrino as a function of
$m_t$ and the VEV ratio. Furthermore, such an assumption would
effectively rule out the existence of the 4th generation of fermions.

We~\footnote{{\normalsize work done in collaboration with {\bf P.~Overmann}
and {\bf E.A.~Paschos}, Dortmund Univ., Germany}} also investigated the
most general framework of the SM with two Higgs doublets such that
no flavor-changing neutral currents (FCNC) occur at the tree level.
Finite 1-loop-induced FCNC (and CP-violating) effects, when confronted
with experimental constraints from the physics of K and B mesons,
provide us with constraints on the values of the dominant Yukawa
couplings of the charged Higgs with the top quark. In the usual, more
restrictive, ``type II'' model, this would imply certain
constraints on the value of the VEV ratio.

\newpage

\noindent{\large {\bf 1.) Flavor Democracy}}

\vspace{0.2cm}

\renewcommand{\baselinestretch}{0.9}

\baselineskip0.7ex

\small\normalsize

We discuss in this part the behavior of different frameworks of the
Standard Model with two Higgs doublets (2HD-SM) at high energies, with
a view to the notion of the so called flavor democracy (FD). Two types
of such models are frequently being used:

a) ``type I''  $[$2HD-SM(I)$]$ -
In this model just one Higgs doublet (say, $H^{(1)}$) couples to all fermions
\begin{equation}
\L^{(I)}_{Yukawa}  =           -  \sum_{i,j=1}^3 \lbrace
D_{ij}^{(q)}( \bar q^{(i)}_L H^{(1)} ) q^{(j)}_{dR} +
U_{ij}^{(q)}( \bar q^{(i)}_L \tilde H^{(1)} ) q^{(j)}_{uR} + \mbox{h.c.}
\rbrace + \cdots \ ,
\end{equation}
where the dots represent the analogous terms for the leptons. The following
notations are used:
\begin{displaymath}
H  =  {H^{+} \choose H^0} \ , \ \tilde H  = i \tau_2 H^{\ast} \ ,
\quad q^{(i)} = {q^{(i)}_u \choose q^{(i)}_d} \ , \quad
q^{(1)} = {u \choose d} \ , \ q^{(2)} = {c \choose s} \ , \
q^{(3)} = {t \choose b} \ ,
\end{displaymath}
and similarly for the leptonic doublets $\ell^{(i)}$ containing Dirac
neutrinos and charged leptons.

This model is very closely related to the minimal SM (MSM), the
only difference in the Yukawa sector being the smaller vacuum expectation
value $\langle (H^0)^{(1)} \rangle_o = v_1/\sqrt{2} < v/\sqrt{2}$ ($v \approx
246.22 GeV$), and hence the correspondingly larger Yukawa coupling parameters.

b) ``type II'' $[$2HD-SM(II)$]$ -
Here, one doublet ($H^{(1)}$) couples to the ``down-type'' right-handed
fermions $f_{dR}$ ($q_{dR}$, $\ell_{dR}$) and is responsible for the
``down-type'' masses ($H^{(1)} \mapsto H^{(d)}$); the other doublet
($H^{(2)} \mapsto H^{(u)}$) couples to the ``up-type'' fermions
$f_{uR}$ ($q_{uR}$, $\ell_{uR}$), being responsible for their masses:
\begin{equation}
\L^{(II)}_{Yukawa}  =           -  \sum_{i,j=1}^3 \lbrace
D_{ij}^{(q)}( \bar q^{(i)}_L H^{(d)} ) q^{(j)}_{dR} +
U_{ij}^{(q)}( \bar q^{(i)}_L \tilde H^{(u)} ) q^{(j)}_{uR} + \mbox{h.c.}
\rbrace + \cdots \ .
\end{equation}
The mass matrices are proportional to the vacuum expectation values (VEVs)
of the Higgses:
$M_u^{(q,\ell)} = v_u U^{(q,\ell)}/\sqrt{2}$,
$M_d^{(q,\ell)} = v_d D^{(q,\ell)}/\sqrt{2}$ , where
\begin{displaymath}
\langle H^{(u)} \rangle_o = \frac{e^{i\xi}}{\sqrt{2}} {0 \choose v_u} \ ,
\qquad
\langle H^{(d)} \rangle_o = \frac{1}{\sqrt{2}} {0 \choose v_d} \ ,
\qquad v_u^2+v_d^2=v^2 (\approx 246^2 GeV^2) \ .
\end{displaymath}
The expressions are written in any flavor basis, i.e.~a basis
in which $f^{(i)}_L$ are $SU(2)_L$-isodoublets.

We note that the 2HD-SM(II), in comparison with the MSM and the closely
related 2HD-SM(I), has essentially {\it different}
(1-loop) renormalization group equations (RGEs) for the
$3 \times 3$
Yukawa matrices
\begin{equation}
16 \pi^2 \frac{dU^{(q)}}{d\ln E} = + \alpha D^{(q)} D^{(q)\dag} U^{(q)}
+ \cdots \ , \qquad \mbox{etc.}
\end{equation}
where $\alpha= 0.5$ in the 2HD-SM(II) ($\alpha=-1.5$ in the 2HD-SM(I) and
MSM). $E$ is here the running energy of probes. It turns out that this
difference in the structure of the RGEs leads to a drastically
different behavior of the corresponding models at high energies
vis-\`a-vis the so called flavor democracy (FD), as will be shown
in the following paragraphs.

We define that a model has the trend to FD (as the energy $E$ increases
toward the Landau pole $\Lb_{pole}$)
iff there exists a flavor basis (i.e.~a basis in which $f_L$ behave as
$SU(2)_L$-isodoublets) such that the Yukawa couplings in the
corresponding ``up-type'' and ``down type'' sectors of quarks (and
leptons) converge to common values and the CKM-mixings converge to zero
\begin{equation}
U^{(q,\ell)}  \to  g^u_{q,\ell} A_{FD} \ , \qquad
D^{(q,\ell)} \to g^d_{q,\ell} A_{FD}
 \ , \qquad V_{ckm} \to \ \mbox{{\large {\bf 1}}} \qquad
(\mbox{as } E \uparrow \Lb_{pole}) \ .
\end{equation}
Here, {\bf 1} is the $3 \times 3$ identity matrix and $A_{FD}$ is the
$3 \times 3$ flavor-democratic matrix
\small
\begin{displaymath}
A_{FD} =
\left[ \begin{array}{ccc}
1 & 1 & 1 \\
1 & 1 & 1 \\
1 & 1 & 1
\end{array} \right] \ .
\end{displaymath}
\normalsize
The motivation behind this notion of FD-trend is the following. We could
interpret the Landau pole $\Lb_{pole}$ (i.e.~the energy where the Yukawa
couplings diverge) as roughly the energy where the SM is replaced by
a new physics ($\Lb_{pole} \sim E_{trans.}$). Then the trend to FD, as
defined above, would represent a signal that the physics beyond the SM
is a strongly interacting physics with almost flavor-blind forces (where
the tiny deviation from FD is provided by some even higher, unknown physics).
However, the entire notion of the trend toward FD at high energies could
also be regarded as independent of any specific motivation.
On the other hand, the notion of FD at {\it low} energies is well
known~\cite{fdlowenergy}.

The FD matrix $A_{FD}$ can be written in the diagonal (``mass'')
basis as the matrix with the diagonal elements $(0,0,3)$. Hence, when
safely neglecting the very light 1st generation masses, we can rewrite
the trend toward FD for the quark sectors (q-q FD) as
\begin{equation}
\frac{m_s}{m_b} \ , \ \frac{m_c}{m_t} \ , \  V_{cb} \ \to \ 0
\qquad (\mbox{as } E \uparrow \Lb_{pole}) \ ,
\end{equation}
and analogously in the leptonic sectors ($\ell$-$\ell$ FD), by replacing
the quark masses by the corresponding masses of charged leptons
and Dirac neutrinos.
We note that already at low energies ($E \sim E_{ew}$), the SM is not
far away from FD.
By simply evolving the 1-loop RGEs for Yukawa couplings (no Higgs
couplings involved) from low energies, where the ``boundary conditions''
are more or less known by experiments, to high energies, we arrive
at the following results:
\begin{displaymath}
\mbox{MSM and 2HD-SM(I)} \ \not\to \ \mbox{FD} \ ;
\qquad \mbox{2HD-SM(II)} \ \to  \ \mbox{FD} \qquad
(\mbox{as } E \uparrow \Lb_{pole}) \ ,
\end{displaymath}
as seen from Figs.~1 and 2.

\vspace{11.cm}
{\small {\bf Figs.~1, 2}: Quark flavor democracy parameters {\normalsize
$m_s/m_b$, $m_c/m_t$} and {\normalsize $V_{cb}$} as functions of the energy
scale $\mu$ ($=E$),
for the case of the minimal SM (left) and the 2HD-SM(II) with
the VEV ratio {\normalsize $v_u/v_d = 0.5$} (right), for various top quark
masses {\normalsize $m_t(m_t)$} (masses of leptons were ignored).}
\vspace{0.3cm}

Figs.~1 and 2 are for the quark sectors (q-q FD), for various
values of $m_t(E=m_t)$ ($ \approx .96 m_t^{phy}$). Fig.~2 is for a
specific ratio of vacuum expectation values (VEVs): $\tan \beta =
v_u/v_d = 0.5$. In the MSM (and the 2HD-SM(I)), it is the ``down-type''
sector and the mixing that lead us away from FD as the energy increases
(cf.~Fig.~1).
However, all the conclusions are the same also for the leptonic sectors
($\ell$-$\ell$ FD), and for all other values of the VEVs. The convergence
toward FD (in the 2HD-SM(II)) is the more striking ($\Lb_{pole}$ lower), the
lower the VEV-ratio $\tan \beta$ and the higher $m_t^{phy}$.
Specifically, for $m_t^{phy}=175GeV$ and $\tan \beta \leq 0.5$,
we have $\Lb_{pole} ( \sim E_{trans.}) \leq 33 TeV$. The related fact
is that choosing $m_t^{phy} \leq 200 GeV$, we obtain $\Lb_{pole}
(\sim E_{trans.}) \stackrel{<}{\sim} E_{Planck}$ for all
$\tan \beta \stackrel{<}{\sim} 1.75$, thus having for a large range of
pertubatively allowed $\tan \beta$
a reasonable framework for condensation mechanism ($H^{(u)} \sim
\langle t \bar t \cdots \rangle$, $H^{(d)} \sim \langle b \bar b +
\cdots \rangle$), {\it unlike} the MSM and the closely related 2HD-SM(I)
(cf.~\cite{bhl}, where the condition $E_{condensation} \sim \Lb_{pole}
\stackrel{<}{\sim} E_{Planck}$ yields $m_t^{phy} \stackrel{>}{\sim}
215 GeV$). Condensation mechanisms leading from a Higgless higher
physics to 2HD-SMs have been seriously investigated~\cite{h2condens}.

Up to this point, the results presented here can be regarded
as formal properties of SMs with various Higgs sectors, formally
independent of any assumptions about the physics beyong the SMs.
However, in the FD-favored 2HD-SM(II), in order to reduce further
the number of high energy Yukawa parameters, we can {\it choose}
to impose the condition of the trend to the so called mixed
quark-lepton FD (q-$\ell$ FD)
\begin{equation}
g^u_q \approx g^u_{\ell} \ , \qquad g^d_q \approx g^d_{\ell}
\qquad \mbox{i.e. } \
\frac{m_{\tau}}{m_b} \ , \ \frac{m_{\nu_{\tau}^D}}{m_t} \ \approx \ 1
\qquad (\mbox{as } E \uparrow \Lb_{pole}) \ .
\end{equation}
\begin{minipage}{8.9cm} \parindent1.cm
Demanding this, and employing the (1-loop) RGEs, we obtain, for
a given $m_t^{phy}$ and $\tan \beta$, the values of the
Dirac mass $m^o_{\nu_{\tau}^D}$
(at $E=1 GeV$) and $\Lb_{pole}$. If employing in addition the usual
see-saw mechanism~\cite{see-saw}, this would give us a good estimate
(upper bound) for the physical neutrino mass
\begin{displaymath}
M_{Majorana} \stackrel{>}{\sim} \Lb_{pole} \ \Rightarrow \
\end{displaymath}
\begin{displaymath}
m^{phy}_{\nu_{\tau}}  \simeq  \frac{(m^o_{\nu_{\tau}^D})^2}{M_{Maj.}} \
\stackrel{<}{\sim} \frac{(m^o_{\nu_{\tau}^D})^2}{\Lb_{pole}} \ (=
(m^{phy}_{\nu_{\tau}})^{u.b.}) \ .
\end{displaymath}
Demanding that this quantity not exceed the experimentally suggested
upper value of $31 MeV$, we obtain an upper bound for $m_t^{phy}$
as a function of $\tan \beta$; similarly, the condition $\Lb_{pole}
< E_{Planck}$ provides us with the lower bound on $m_t^{phy}$ (Fig.~3).
For $m_t^{phy}=(175 \pm 20) GeV$, we obtain from Fig.~3:
$0.53 < \tan \beta < 2.1$. These results, however, are obtained under
the assumption of the imposed q-$\ell$ FD at $E \simeq \Lb_{pole}$ (in the
FD-favored 2HD-SM(II)).

Most of the results presented here up to this point have been published
in refs.\cite{csk}.
\end{minipage}

When imposing the q-$\ell$ FD (at $E \simeq \Lb_{pole}$) in the 2HD-SM(II),
we can ask ourselves whether such a framework would allow for the
existence of the (heavy) 4th generation of fermions ($t^{\prime},
b^{\prime}$), ($\nu^D_{\tau^{\prime}}, \tau^{\prime}$). In this case,
the q-$\ell$ FD would mean:
$m_{t^{\prime}} \approx m_{\nu^D_{\tau^{\prime}}} \ , \
m_{b^{\prime}} \approx m_{\tau^{\prime}} \qquad
\mbox{at } E \approx \Lb_{pole} \ .$
Similarly as before, for any chosen $m^{phy}_{t^{\prime}}$ and
$m^{phy}_{b^{\prime}}$
(and for given 3rd generation masses $m^{phy}_t$, $m^{phy}_b$ and
$m^{phy}_{\tau}$), we obtain by employing the 1-loop RGEs the correspondence
\begin{displaymath}
m^{phy}_{t^{\prime}} \ , \ m^{phy}_{b^{\prime}} \ \mapsto \ \Lb_{pole} \ ,
\ (m^{phy}_{\nu_{\tau^{\prime}}})^{u.b.} \ , \  m^{phy}_{\tau^{\prime}} \ .
\end{displaymath}
The correspondence is virtually independent of the Yukawa coupling of
the 3rd generation neutrino $\nu^D_{\tau}$ (at $E = 1GeV$).
We applied see-saw, as before, to obtain the upper bound estimate
$(m^{phy}_{\nu_{\tau^{\prime}}})^{u.b.}$. However, at least 4 additional
experimentally suggested constraints have to be imposed: \\
\vspace{0.2cm}
\centerline{
$1.) \ m^{phy}_{t^{\prime}},m^{phy}_{b^{\prime}} > m^{phy}_t \ ; \qquad$
$2.) \ 40GeV \leq m^{phy}_{\nu_{\tau^{\prime}}} \ (
\stackrel{<}{\sim} (m^o_{\nu^D_{\tau^{\prime}}})^2/\Lb_{pole})$ ;} \\
\vspace{0.1cm}
\centerline{
$3.) \ \mbox{ the SM is perturbative (say:} \
 m^{phy}_{t^{\prime}}, m^{phy}_{b^{\prime}} \stackrel{<}{\sim} 0.5 \Lb_{pole})
\ ; \quad $
\vspace{0.1cm}
$4.) \ (\triangle \rho )_{\mbox{{\scriptsize from heavy fermions}}} \leq
0.0076$.
}\\
\vspace{0.1cm}
The fourth constraint is from an essentially model-independent analysis of
the LEP data~\cite{delrho}. It turns out that the four constraints above
can be satisfied only if $m_t^{phy} < 155 GeV$. Hence, in the light of the
recent CDF measurement of $m_t^{phy}$ ($\simeq (174 \pm 20)GeV$), we
conclude that the existence of the 4th generation is practically ruled
out within this framework (i.e.~in the FD-favored 2HD-SM(II), with
see-saw, and q-$\ell$ FD at $\Lb_{pole}$)~\cite{cskprep}.

\vspace{0.3cm}

\noindent{\large {\bf 2.) Suppression of the flavor-changing
neutral currents in the 2HD-SM(II)}}

\vspace{0.2cm}

Experiments show that the flavor-changing neutral currents (FCNC) are
very suppressed
\begin{displaymath}
\frac{\Gamma(K^0_L \to \mu^+\mu^-)}{\Gamma(K^0_L \to all)} \simeq
(7.3 \pm 0.4)10^{-9} \ , \qquad |m_{B^0_d}-m_{\bar{B^0_d}} | \simeq
(3.6 \pm 0.7) 10^{-10}MeV \ ,
\end{displaymath}
\begin{displaymath}
 |m_{K_L}-m_{K_S}| \simeq (3.52 \pm 0.02) 10^{-12}MeV \ , \ \mbox{etc.}
\end{displaymath}
Therefore, the usual conditions imposed in various SMs are:
a) {\it no} FCNC at the tree level, and
b) the 1-loop-induced FCNC are sufficiently suppressed.

The conditions for the 1-loop FCNC suppression for gauge boson loops,
i.e.~the allowed representation contents of fermions, have been
investigated some time ago~\cite{paschos}. In addition, Glashow and
Weinberg~\cite{paschos} proposed for the Higgs sector the MSM
(i.e.~one Higgs doublet) and ``type I'' and/or ``type II''
2HD-SM (two Higgs doublets; cf. eqs.~(1) and (2)) - as models which,
in a ``natural'' way, have no FCNC in the Yukawa couplings
at the tree level.

However, recently Y.-L.~Wu~\cite{wu} has pointed out the most general
framework of the 2HD-SMs not having FCNC at the tree level in the
Yukawa sector (``type III''). The Yukawa interactions in this
framework, in any flavor basis, have the most general form
\begin{eqnarray}
\L^{(III)}_{Y.} & = & - \sum_{i,j=1}^3 \lbrace
D_{ij}^{(1)}( \bar q^{(i)}_L H^{(1)} ) q^{(j)}_{dR} +
D_{ij}^{(2)}( \bar q^{(i)}_L H^{(2)} ) q^{(j)}_{dR} +
    \nonumber\\
& & + U_{ij}^{(1)}( \bar q^{(i)}_L \tilde H^{(1)} ) q^{(j)}_{uR} +
U_{ij}^{(2)}( \bar q^{(i)}_L \tilde H^{(2)} ) q^{(j)}_{uR} + h.c.
\rbrace
+ \lbrace  \ \bar \ell\mbox{H}\ell\mbox{-terms} \ \rbrace \ ,
\end{eqnarray}
where, however, the $3 \times 3$-Yukawa matrices $D^{(1)}$, $D^{(2)}$,
$U^{(1)}$ and $U^{(2)}$ are now such that in the mass basis of quarks
(fermions) they are all {\it simultaneously} diagonal.
We note that ``type I'' and ``type II'' models
(eqs.(1) and (2)) are special cases of this 2HD-SM(III). The corresponding
charged-current part of the quarks can then be deduced
\small
\begin{equation}
\L^{(III)cc}_{Y.} = H^{(+)} \lbrack - \overline{u_L}V D d_R +
\overline{u_R} U^{\dag}
V d_L \rbrack + H^{(-)} \lbrack \mbox{ h.c. } \rbrack \ .
\end{equation}
\normalsize
This expression is in the unitary gauge and in the physical bases
(for fermions and the charged Higgses $H^{(\pm)}$). $V$ is the CKM-matrix,
$u^T = (u,c,t)$, $d^T = (d,s,b)$; $U$ and $D$ are specific linear
combinations of the diagonal Yukawa matrices $U^{(j)}$ and $D^{(j)}$
\small
\begin{displaymath}
U  =  - \sin \beta U^{(1)} + \cos \beta e^{-i \xi} U^{(2)} \ ,
\qquad
D  =  - \sin \beta D^{(1)} + \cos \beta e^{+i \xi} D^{(2)} \ ,
\end{displaymath}
\normalsize
where $\tan \beta$ is the ratio of the absolute values of the
VEVs ($\tan \beta = v_2/v_1$), the mass basis is used,
and $\xi$ is the CP-violating phase between the VEVs.
On the other hand, the (diagonal) mass matrices are obtained from the
corresponding orthonormal linear combinations
\small
\begin{displaymath}
\frac{\sqrt{2}}{v}M_u  =
 \cos \beta U^{(1)} + \sin \beta e^{-i \xi} U^{(2)} \ ,
\qquad
\frac{\sqrt{2}}{v}M_d  =
 \cos \beta D^{(1)} + \sin \beta e^{+i \xi} D^{(2)} \ .
\end{displaymath}
\normalsize
Therefore, assuming that no peculiar cancelations occur, $U_{33}$
($\sim (m^{phy}_t/v)$) is by far the most dominant Yukawa coupling
in $\L^{(III)cc}_{Y.}$.

{\noindent
\begin{minipage}{8.9cm} \parindent1.cm
Since the $H^{(\pm)}$-exchanges influence the $\bar{B^0_d}$-$B^0_d$ and
$\bar{K^0}$-$K^0$ mixing, the experimental values of
$|m_{B^0}-m_{\bar{B^0}}|$, $|m_{K_L}-m_{K_S}|$ and of the
CP-violating parameters $\varepsilon_K$ and $\varepsilon_K^{\prime}$
would provide us
with restrictions on the (dominant) $U_{33}$ coupling strength
of $H^{(\pm)}$ to quarks. The dominant graphs contributing to these
mass differences are the W-W, H-W and H-H exchange box diagrams
(cf.~Fig.~4). The resulting effective 4-fermion couplings are
\end{minipage}}
\small
\begin{eqnarray}
\L_{eff}  (= \L_{eff}^{WW} + \L_{eff}^{HW} + \L_{eff}^{HH})
 & \simeq & \A^K [\overline{d(x)^a} \gamma^{\mu}
(\frac{1-\gamma_5}{2})s(x)^a]^2 \quad (\mbox{for }  K^0 - \bar{K^0})
\nonumber\\
&  \simeq & \A^B [\overline{b(x)^a} \gamma^{\mu}
(\frac{1-\gamma_5}{2})d(x)^a]^2
\quad (\mbox{for } B_d^0 - \bar{B_d^0}) \ ,
\end{eqnarray}
\begin{displaymath}
\mbox{where:} \qquad \A  =  \A_{WW} + \A_{HW} + \A_{HH} \ ,
\qquad
\A_{HW}  =  \frac{G_F}{4 \sqrt{2} \pi^2} \J(\frac{M_{H^-}}{M_W},
\frac{m_t}{M_W}) (\frac{m_t}{M_W})^2 \zeta_t^2 |U_{33}|^2 + \cdots \ ,
\end{displaymath}
\begin{displaymath}
\A_{HH}  =  - \frac{1}{64 \pi^2 M_{H^-}^2} \I(\frac{m_t}{M_{H^-}})
\zeta_t^2 |U_{33}|^4 + \cdots \ ,
\qquad
\A_{WW}  =  \frac{G_F^2 M_W^2}{4 \pi^2} \lbrace \E(\frac{m_c}{M_W})
\zeta_c^2 + \cdots \rbrace \ .
\end{displaymath}
We denoted here
\begin{equation}
\zeta_t  =  V^{\ast}_{td} V_{ts} \ (\mbox{for } K^0-\bar{K^0}) \ ,
 \qquad \zeta_t = V_{td}V^{\ast}_{tb} \ (\mbox{for } \bar{B^0}-B^0) \ ,
\end{equation}
\normalsize
and analogously for $\zeta_c$. $\J$, $\I$ and $\E
$ are dimensionless
and slowly varying functions ($\sim {\cal O}(1)$).
Such effects, for the specific ``type II'' model, have been investigated
by several authors~\cite{gg}; here they are studied within the more
general framework of the ``type III'' model. The relation of these
amplitudes to the experimental inputs of K and B physics is
\small
\begin{eqnarray}
\sqrt{2} (\triangle M_{K^0_L-K^0_S}) (|\varepsilon_K| +
0.05 \frac{\varepsilon^{\prime}_K}{\varepsilon_K}) & \simeq &
- \frac{1}{2 M_{K^0}}
 \mbox{Im} \langle K^0 | \L_{eff}(x=0) | \bar{K^0} \rangle
\nonumber\\
& = & - \frac{1}{2 M_{K^0}} \A^K
\mbox{Im} \langle K^0 | [\overline{d^a} \gamma^{\mu}
(\frac{1-\gamma_5}{2})s^a]^2 | \bar{K^0} \rangle \ ,
\end{eqnarray}
\begin{equation}
M_{B^0} \triangle M_{B^0-\bar{B^0}} \simeq
| \langle \bar{B^0_d} | \L_{eff}(x=0) | B^0_d \rangle | =
\A^B | \langle \overline{B^0_d} | [\bar{b^a} \gamma^{\mu}
(\frac{1-\gamma_5}{2})d^a ]^2 | B^0_d \rangle | \ ,
\end{equation}
\normalsize
where the normalization conventions are: $\langle P^0|P^0 \rangle =
Vol \ast 2 M_{P^0}$ ($P^0=K^0, B^0, \ldots$; $Vol$ is the
3-dimensional volume of the space).
While the box diagram amplitudes $\A^K$, $\A^B$ represent the dominant
perturbative contributions in the above relations, the matrix elements
Im$\langle K^0 | \cdots | \bar{K^0} \rangle$ and
$\langle \bar{B^0_d} | \cdots | B^0_d \rangle$
represent the hadronic (non-perturbative) effects. Experiments provide us
with the following values for the $\triangle M$'s and the CP-violating
parameters $\varepsilon_K$ and $\varepsilon^{\prime}_K$:
\small
\begin{eqnarray}
\sqrt{2} (\triangle M_{K^0_L-K^0_S}) (|\varepsilon_K| +
0.05 \frac{\varepsilon^{\prime}_K}{\varepsilon_K}) & \simeq &
1.16 (1 \pm 0.03 ) 10^{-17} GeV \ ,
\nonumber\\
\triangle M_{B^0-\bar{B^0}} & \simeq & (3.6 \pm 0.7) 10^{-13} GeV \ .
\end{eqnarray}
\normalsize
For the hadronic matrix elements, we have the following (theoretical)
uncertainties:
\small
\begin{displaymath}
\langle K^0 | [\bar{d^a} \gamma^{\mu} (\frac{1-\gamma_5}{2})s^a]^2 |
\bar{K^0} \rangle  =  \frac{2}{3} F^2_K B_K M_{K^0}^2 \ , \qquad
0.6 \stackrel{<}{\sim} B_K \stackrel{<}{\sim} 0.9 \ .
\end{displaymath}
\begin{displaymath}
\langle \bar{B^0_d} | [\bar{b^a} \gamma^{\mu}(\frac{1-\gamma_5}{2})
d^a]^2 | B^0_d \rangle  =  \frac{2}{3} F_B^2 B_B M_{B^0}^2 \ , \qquad
0.15 GeV \stackrel{<}{\sim} F_B \sqrt{B_B} \stackrel{<}{\sim} 0.2 GeV \ .
\end{displaymath}
\normalsize
In addition, we have many uncertainties also in the CKM-matrix $V$
(in $\L^{(III)cc}_{Y.}$). We chose Maiani's notation, fixing the real angles
$\theta_{12}$, $\theta_{13}$, $\theta_{23}$ at their experimental
average and allowing the CP-violating phase $\delta^{\prime}$ in $V$
to be free. Then, for a chosen $m_t^{phy}$ and $M_{H^-}$, we obtain in
the plane $\delta^{\prime}$ vs.~$|U_{33}|$ a stripe allowed by the
hadronic uncertainties of the $\bar{B^0}$-$B^0$ mixing, and another stripe
allowed by the hadronic uncertainties of the $K^0$-$\bar{K^0}$ mixing.
For $m^{phy}_t=175 GeV$ and $M_{H^-}=1TeV$, these stripes are depicted
in Fig.~5, as well as their overlap (we use instead of $|U_{33}|$ the
``normalized'' parameter $Z=|U_{33}|/(\sqrt{2}m^{phy}_t/v)$). We included
the leading (known) QCD corrections to the W-W exchange box diagrams, by
making the following replacements in the amplitude $\A^K_{WW}$:
$\zeta_c^2 \mapsto 0.81 \zeta_c^2$, $\ \zeta_t^2 \mapsto 0.59 \zeta_t^2$,
$\ \zeta_t \zeta_c \mapsto 0.37 \zeta_t \zeta_c$; and in $\A^B_{WW}$:
$\zeta_t^2 \mapsto 0.8 \zeta_t^2$.
This changes the curves and the overlap quite substantially in the region
$|U_{33}| \approx 0$. When decreasing $M_{H^-}$ to $200 GeV$, the features
remain unchanged, but the upper bound for $|U_{33}|$ becomes smaller
(Fig.~6). We note that it is the $\bar{B^0}$-$B^0$ mixing that provides
us (for a given $m^{phy}_t$ and $M_{H^-}$) with a rather restrictive
upper bound on the dominant Yukawa coupling $|U_{33}|$. For heavier
top quarks, this upper bound becomes even more restrictive (lower).
Furthermore,
in the special ``type II'' case, the horizontal axis in Figs.~5 and 6 should
be interpreted as $\cot \beta$ ($= v_d/v_u$). In this case, Figs.~5 and 6
would give us lower bounds for $\tan \beta$.

\vspace{8cm}
{\small {\bf Figs.~5 and 6}: The regions in the {\normalsize
$\delta^{\prime}$}
vs.~{\normalsize Z ($=|U_{33}|/(2m^{phy}_t/v)$)} parameter plane as allowed by
the
hadronic uncertainties for the case of $\bar{B^0_d}$-$B^0_d$ mixing
(thick curves) and $\bar{K^0}$-$K^0$ mixing (thinner curves), for
the case of the charged Higgs mass $M_{H^-}=$ 1TeV (left) and
$M_{H^-}=$0.2TeV (right). The overlap regions are also denoted.
We took {\normalsize $m_t^{phy}=$}175GeV, and took
into account the leading QCD corrections for the W-W box exchange diagrams.}
\vspace{0.3cm}

These rather preliminary calculations call for further improvements:
\begin{itemize}
\item the QCD corrections should also be included in the H-W and
H-H exchange box diagrams (Fig.~4), thus improving Figs.~5 and 6 in the
regime of large $|U_{33}|$;

\item especially the experimentally uncertain $\theta_{13}$ angle
of the CKM-matrix (in Maiani's notation) should be varied within the
allowed range, thus giving us somewhat modified figures;

\item similar calculations should be performed also for the
``down-type'' hadrons, e.g.~for the $D^0$-$\bar{D^0}$ mixing,
possibly giving us clues to restrictions on the dominant ``down-type''
coupling $D_{33}$ of charged $H^+$ to $b_R$ quark.

\end{itemize}

\vspace{0.3cm}

\noindent{\large {\bf Conclusions}}

\vspace{0.1cm}
\baselineskip0.6ex

In general, standard models with two (or more) Higgs
doublets (2HD-SMs) have a rich physics, especially
concerning the CP violation~\cite{cpv}. However, there are
also several other features that make these models attractive and
very different from the minimal SM (MSM):
\begin{enumerate}
\item The 2HD-SM(II) has a clear and consistent convergence to
flavor democracy as $E \uparrow \Lb_{pole}$ ({\it unlike} the
MSM and the closely related 2HD-SM(I)), thus giving us a signal of a
possible new and (almost) flavor-blind strongly interacting physics
at the energies beyond the Landau pole:
$E \stackrel{>}{\sim} \Lb_{pole}$ ($\stackrel{>}{\sim} 1TeV$).

\item For $\Lb_{pole}$ ($\sim E_{trans.}$) being smaller than
$E_{Planck}$, it is possible to have $m_t^{phy} < 200 GeV$ in the
2HD-SM(II) ({\it not} in the MSM and 2HD-SM(I)) $\Longrightarrow$
condensation scenarios leading to the 2HD-SM(II) (and to approximate
flavor democracy at $E_{trans.} \sim E_{condens.} \sim \Lb_{pole}$)
may be promising.

\item The {\it imposed} mixed ``lepton-quark'' flavor democracy
at $E \simeq \Lb_{pole}$ in the 2HD-SM(II) leads to restricted values of
the ratio of the VEVs ($0.53 < v_u/v_d < 2.1$, for $m_t^{phy}=
(175 \pm 20) GeV$), and furthermore, would make the existence of the 4th
generation of fermions very unlikely.

\item 2HD-SM(III) is the most general 2HD-SM with {\it no}
Higgs-mediated flavor-changing neutral currents (FCNC) at the tree level,
and is a promising framework for investigations; experimental FCNC
(and CP violation) constraints, when confronted with 1-loop and higher
loop effects, may severely restrict the parameter space of the model.

\end{enumerate}

\newpage

\small \normalsize

\begin{minipage}{7cm}
{\small {\bf Fig.~3}: Bounds on $m^{phy}_t$ as function of the VEV ratio
$v_u/v_d$ in the 2HD-SM(II); assumed: q-$\ell$ FD at $\Lb_{pole}$, and
see-saw.}
\end{minipage}

\vspace{1.cm}

\begin{minipage}{7cm}
{\small {\bf Fig.~4}: W-W, H-W and H-H exchange box diagrams for
$\bar{P^0}$-$P^0$ mixing ($P^0$= $B^0$, $K^0$).}
\end{minipage}

\small\normalsize

\small

\begin{thebibliography}{99}

\bibitem{fdlowenergy}
Y.~Nambu, Proc.~of XI Warsaw Symposium on High Energy Physics (1988);
H.~Harari, H.~Haut and J.~Wegers, Phys.~Lett.~{\bf B78} (1978) 459;
P.~Kaus and S.~Meshkov, Phys.~Rev.~{\bf D42} (1990) 1863;
F.~Cuypers and C.S.~Kim, {\it ibid.}~{\bf B254} (1991) 462;
H.~Fritzsch and J.~Plankl, Phys.~Lett.~{\bf B237} (1990) 451

\bibitem{bhl}
W.A.~Bardeen, C.T.Hill and M.~Lindner, Phys.~Rev.~{\bf D41} (1990) 1647

\bibitem{h2condens}
M.~Suzuki, Phys.~Rev.~{\bf D41} (1990) 3457; M.~Harada and N.~Kitazawa,
Phys.~Lett.~{\bf B257} (1991) 383

\bibitem{see-saw}
T.~Yanagida, Proc.~Workshop on Unified Theory and Baryon
Number of the Universe (KEK, Japan, 1979);
M.~Gell-Mann, P.~Ramond and R.~Slansky, in ``Supergravity'', ed.
P.~Van Nieuwenhuizen and D.Z.~Freedman (North-Holland, Amsterdam, 1979)

\bibitem{csk}
G.~Cveti\v c and C.S.~Kim, Nucl.~Phys.~{\bf B407} (1993) 290,
Mod.~Phys.~Lett.~{\bf A9} (1994) 289

\bibitem{delrho}
A.~Blondel and C.~Verzegnassi, Phys.~Lett.~{\bf B311} (1993) 346,
and private communication

\bibitem{cskprep}
G.~Cveti\v c and C.S.~Kim, Dortmund preprint DO-TH-94/09

\bibitem{paschos}
E.A.~Paschos, Phys.~Rev.~{\bf D15} (1977) 1966;
S.L.~Glashow and S.~Weinberg, Phys.~Rev.~{\bf D15} (1977) 1958;

\bibitem{wu}
Y.-L.~Wu, Carnegie Mellon preprint CMU-HEP 93-19

\bibitem{gg}
J.F.~Gunion and B.~Grz\c adkowski, Phys.~Lett.~{\bf B243} (1990) 301;
A.J.~Buras, P.~Krawczyk, M.E.~Lautenbacher and
C.~Salazar, Nucl.~Phys.~{\bf B337} (1990) 284

\bibitem{cpv}
B.~Grz\c adkowski, Proc.~of the XXVIIIth Rencontre de Moriond
(``'93 Electroweak Interactions and Unified Theories''), and
refs.~therein; G.~Cveti\v c, {\it ibid};
Y.-L.~Wu, preprint CMU-HEP 94-02

\end{thebibliography}
\end{document}